\documentclass[twocolumn]{revtex4}
\usepackage{mathrsfs}            
\usepackage{amssymb}
\usepackage{amsmath}             
\usepackage[dvips]{graphicx}
\usepackage{wasysym}

\usepackage{texdraw}    
\input{txdtools}        
\usepackage{ifthen}     

\newcommand{\qdrawdim}{pt}
\newcommand{\qlinewidth}{.5}
\newlength{\defaultunitlength}
\newcommand{\qcircuitcalc}[3]{\drawdim{\qdrawdim} \linewd{\qlinewidth}
                              \realmult{#1}{1}          \step
                              \realadd {#2}{1}          \slots
                              \realmult{\step}{.5}      \halfstep
                              \realmult{\step}{.7}      \ministep
                              \realmult{\ministep}{.5}  \halfministep
                              \realmult{\ministep}{.1}  \margin
                              \realmult{\margin}{2.}    \doublemargin
                              \realmult{\ministep}{.3}  \ballradius
                              \realmult{\step}{.1}      \dotradius
                              \realmult{\dotradius}{1.} \multilineskip
                              \realmult{\step}{\slots}  \tot
                              \realmult{#3}{1}          \text
                              \realadd {\tot}{\text}    \ttot
                              \realadd {0}{1}           \theslot
                              \realmult{1}{1}           \boxfill
                              \realmult{\qlinewidth}{2} \erasewidth
                              \realadd {\dotradius}{-\erasewidth} \yesdotradius
                              \setlength{\defaultunitlength}{\unitlength}
                             }
\newcounter{line_counter}
\newenvironment{qcircuit}[4]{\begin{tabular}[c]{c}\begin{texdraw}
                             \qcircuitcalc{#1}{#2}{#4}
                             \setcounter{line_counter}{#3}
                             \whiledo{\value{line_counter} > 0}
                             {\addtocounter{line_counter}{-1}
                              \qline{\value{line_counter}}
                             }
                            }
                            {\end{texdraw}\end{tabular}}

\newcommand{\qsame}[1][1]{\realadd{\theslot}{-#1}\theslot}
\newcommand{\qskip}[1][1]{\realadd{\theslot}{#1}\theslot}
\newcommand{\qcalcpos}[1]{\realmult{#1}{\step}          \vert
                          \realmult{\step}{4}           \vertoffset
                          \realadd{\vert}{\vertoffset}  \vert 
                          \realmult{\theslot}{\step}    \hori
                          \realadd {\hori}{\text}       \thori}
\newcommand{\qboxspace}{\rmove({-\halfministep} {-\halfministep})
                        \rmove({\ministep} 0)  \rmove(0 {\ministep})
                        \rmove({-\ministep} 0) \rmove(0 {-\ministep}) 
                        \rmove({\halfministep} {\halfministep})}
\newlength{\qtextlength}
\newcommand{\qtextL}[1]{\textref h:L v:C \htext{#1} \drawdim{sp}
                        \settowidth{\qtextlength}{#1}
                        \rmove({\number\qtextlength} 0) \drawdim{\qdrawdim}}
\newcommand{\qtextR}[1]{\textref h:R v:C \htext{#1} \drawdim{sp}
                        \settowidth{\qtextlength}{#1}
                        \rmove({-\number\qtextlength} 0) \drawdim{\qdrawdim}}
\newcounter{domultilines}
\newcommand{\qgenerateline}[6]{\qcalcpos{#1} \move({\text} {\vert})
                               \realmult{#2}{\step}\frompatch
                               \realmult{#3}{\step}\topatch
                               \realadd{\topatch}{-\frompatch}\topatch
                               \setcounter{domultilines}{#5}
                               \realadd{#5}{-1}\vshift
                               \realmult{\multilineskip}{\vshift}\vshift
                               \realmult{\vshift}{0.5}\vshift
                               \rmove({\frompatch} {\vshift}) 
                               \whiledo{\value{domultilines} > 0}
                                    {\addtocounter{domultilines}{-1}
                                     \ifthenelse{\equal{#6}{del}}
                                        {\linewd {\erasewidth} \setgray 1
                                         \rlvec({\topatch} 0)
                                         \linewd {\qlinewidth} \setgray 0
                                         \rmove({-\topatch} 0)}{}
                                     \lpatt(#4) \rlvec({\topatch} 0)
                                     \lpatt() \rmove({-\topatch} 0)
                                     \rmove(0 {-\multilineskip})
                                    }
                               \rmove(0 {\vshift}) \rmove(0 {\multilineskip})
                               }
\newcommand{\qmultiline}[2]{\qgenerateline{#1}{0}{\slots}{}{#2}{del}}
\newcommand{\qline}[1]{\qgenerateline{#1}{0}{\slots}{}{1}{} \qboxspace}

\newlength{\boxwidth}
\newlength{\boxheight}
\newlength{\defaultwidth}
\newlength{\defaultheight}
\newlength{\halfboxwidth}
\newlength{\halfboxheight}
\newlength{\halfheightexcess}
\newcommand{\qgenericbox}[5]{\qcalcpos{#1} \move({\thori} {\vert})
                             \realadd{#2}{-#1} \thelinerange
                             \realmult{\thelinerange}{\step} \linedistance
                             \setlength{\defaultheight}{\linedistance\qdrawdim}
                             \advance\defaultheight by \ministep\qdrawdim
                             \setlength{\defaultwidth}{\ministep\qdrawdim}
                             \setlength{\boxwidth}{#3}
                             \advance\boxwidth by \doublemargin\qdrawdim
                             \setlength{\boxheight}{#4}
                             \advance\boxheight by \doublemargin\qdrawdim
                             \ifthenelse{\lengthtest{\boxwidth<\defaultwidth}}
                                        {\setlength{\boxwidth}{\defaultwidth}}{}
                             \ifthenelse{\lengthtest{\boxheight<\defaultheight}}
                                        {\setlength{\boxheight}{\defaultheight}}{}
                             \setlength{\halfboxwidth}{.5\boxwidth}
                             \setlength{\halfboxheight}{.5\boxheight}
                             \setlength{\halfheightexcess}{-\linedistance\qdrawdim}
                             \advance\halfheightexcess by \boxheight
                             \divide\halfheightexcess by 2
                             \drawdim{sp}
                             \setgray #5 \rmove({-\number\halfboxwidth} {-\number\halfheightexcess})
                             \rlvec({\number\boxwidth} 0) \rlvec(0 {\number\boxheight})
                             \rlvec({-\number\boxwidth} 0) \rlvec(0 {-\number\boxheight})
                             \lfill f:{\boxfill} \setgray 0
                             \rmove({\number\halfboxwidth} {\number\halfboxheight})
                             \drawdim{\qdrawdim}
                            }
\newlength{\qlabelwidth}
\newlength{\qlabelheight}
\newcommand{\qbox}[3]{\settowidth{\qlabelwidth}{#3}
                      \settoheight{\qlabelheight}{#3}
                      \qgenericbox{#1}{#2}{\qlabelwidth}{\qlabelheight}{0}
                      \textref h:C v:C \htext{#3}}

\newcommand{\qsingle}[2]{\qbox{#1}{#1}{#2} \qskip}
\newcommand{\qmultiple}[3]{\qbox{#1}{#2}{#3} \qskip}
\newcommand{\qmeasure}[1]{\qcalcpos{#1} \move({\thori} {\vert})
			  \linewd{\erasewidth} \setgray 1
			  \lvec({\ttot} {\vert})
                          \linewd{\qlinewidth} \setgray 0
                          \qgenericbox{#1}{#1}{0\qdrawdim}{0\qdrawdim}{0}
                          \textref h:C v:C \rmove(0 {-\halfministep})
                          \larc r:{\halfministep} sd:40 ed:140
                          \rmove(0 {\dotradius}) \arrowheadtype t:
                          \arrowheadsize l:{\dotradius} w:{\dotradius}
                          \realadd{\halfministep}{\dotradius}\arrowheight
                          \ravec({\dotradius} {\arrowheight}) 
                          \move({\thori} {\vert}) \qskip }

\newcommand{\qcalcbrace}[2]{\qcalcpos{#1} \realmult{\vert}{1}\bcenter
                            \qcalcpos{#2} \realadd{\vert}{\bcenter}\bcenter
                            \realadd{#2}{-#1}\bheight
                            \realadd{\bheight}{1}\bheight
                            \realmult{\bheight}{\step}  \bheight
                            \realmult{\bcenter}{.5}     \bcenter
                            \realmult{\bheight}{.5}     \bheight
                           }
\newcommand{\qbrace}[5]{\qcalcbrace{#1}{#2}
                        \move(0 {\bcenter}) \textref h:L v:C \htext{#3}
                        \realmult{#5}{\text} \hcenter
                        \move({\hcenter} {\bcenter})
                        \qtextR{#4~\makebox[\ministep pt]%
                                 {$\left\{\rule{0pt}{\bheight pt}\right.$}}}
\newcommand{\qendbrace}[3]{\qcalcbrace{#1}{#2} \move({\ttot} {\bcenter})
                           \qtextL{\makebox[\ministep pt]%
                                {$\left.\rule{0pt}{\bheight pt}\right\}$}~#3}}

\newcommand{\qeraseline}[1]{\qcalcpos{#1} \move({\text} {\vert})
                            \linewd{\erasewidth} \setgray 1
                            \rlvec({\tot} 0)
                            \linewd{\qlinewidth} \setgray 0}
\newcommand{\qnogate}[1]{\qcalcpos{#1} \move({\thori} {\vert})
                         \fcir f:0 r:{\qlinewidth}
                         \rmove(0 {-\halfministep})
                         \fcir f:0 r:{\qlinewidth}
                         \rmove(0 {\ministep})
                         \fcir f:0 r:{\qlinewidth}}
\newcommand{\qparenthesis}[2]{\qcalcbrace{#1}{#2}
                              \qcalcpos{#1}
                              \move({\thori} {\bcenter})
			      \rmove({\halfministep} 0)
			      \qtextR{\makebox[\ministep pt]%
                                     {$\left(\rule{0pt}{\bheight pt}\right.$}}
			      \qskip}
\newcommand{\qendparenthesis}[3]{\qcalcbrace{#1}{#2}
                                 \qcalcpos{#1}
                                 \move({\thori} {\bcenter})
			         \rmove({\halfministep} 0)
			         \qtextR{\makebox[\ministep pt]%
                                        {$\left.\rule{0pt}{\bheight pt}\right)_{\lefteqn{#3}}$}}
			         \qskip}
\newlength{\ministeplength}
\newcommand{\qtrashcan}[1]{\qcalcpos{#1} \move({\thori} {\vert})
			   \linewd{\erasewidth} \setgray 1
			   \lvec({\ttot} {\vert})
                           \linewd{\qlinewidth} \setgray 0
			   \move({\thori} {\vert})
			   \rmove({-\halfministep} {-\halfministep})
			   \realdiv{\ministep}{16} \trashunit
			   \setlength{\ministeplength}{\trashunit\qdrawdim}
			   \drawdim{\ministeplength}
                           \bsegment
			     \move(4 0) \lvec(12 0) \lvec(13 1)
			     \lvec(13 13) \lvec(3 13) \lvec(3 1)
			     \lvec(4 0) \lfill f:1
			     \move(5 0) \lvec(5 13)
			     \move(7 0) \lvec(7 13)
			     \move(9 0) \lvec(9 13)
			     \move(11 0) \lvec(11 13)
			     \move(2 13) \lvec(14 13) \lvec(14 14)
			     \lvec(13 15) \lvec(3 15) \lvec(2 14)
			     \lvec(2 13) \lfill f:1
			     \move(6 15) \lvec(7 16)
			     \lvec(9 16) \lvec(10 15)
			   \esegment
			   \drawdim{\qdrawdim}
                           \move({\thori} {\vert}) \qskip }
\newcommand{\ud}{\mathrm{d}}
\newcommand{\Tr}{\mathrm{Tr}}
\newcommand{\nth}[1]{$#1^{\mbox{\scriptsize th}}$}

\newcommand{\Id}{\ensuremath{\mbox{I\hspace{-.2em}I}}}

\newcommand{\abs}[1]{\ensuremath{\left|#1\right|}}
\newcommand{\ket}[1]{\ensuremath{\left|#1\right\rangle}}
\newcommand{\bra}[1]{\ensuremath{\left\langle#1\right|}}
\newcommand{\mean}[1]{\ensuremath{\left\langle #1 \right\rangle}}

\newcommand{\ketbra}[2]{\ensuremath{\ket{#1}\!\bra{#2}}}

\newlength{\barheight}
\newcommand{\Sandwich}[3]{\settoheight{\barheight}{${\displaystyle #1}$ ${\displaystyle #2}$ ${\displaystyle #3}$}\left\langle#1\left|\rule{0pt}{\barheight}#2\right|#3\right\rangle}   
\newcommand{\Prob}[1]{\ensuremath{\mathrm{Pr}\left(#1\right)}}

\begin{document}
 
\title{Efficient error characterization in Quantum Information Processing}
\author{Benjamin L\'evi$^1$, Cecilia C. L\'opez$^1$, Joseph Emerson$^2$, D. G. Cory$^1$} 
\affiliation{(1): Department of Nuclear Science and Engineering, MIT, Cambridge, MA 02139, USA} 
\affiliation{(2): Department of Applied Mathematics and Institute for Quantum Computing, University of Waterloo, Waterloo, ON N2L 3G1, Canada}

\begin{abstract}
We describe how to use the fidelity decay as a tool to characterize the errors affecting a quantum information processor through a noise generator $G_{\tau}$.
For weak noise, the initial decay rate of the fidelity proves to be a simple way to measure the magnitude of the different terms in $G_{\tau}$.
When the generator has only terms associated with few-body couplings, our proposal is scalable. We present the explicit protocol for estimating the magnitude 
of the noise generators when the noise consists of only one and two-body terms, and describe a method for measuring the parameters of more general noise models.
The protocol focuses on obtaining the magnitude with which these terms affect the system during a time step of length $\tau$; 
measurement of this information has critical implications for assesing the scalability of fault-tolerant quantum computation in any physical setup.
\end{abstract}
\date{\today}

\maketitle


\section{Introduction}

One of the biggest challenges in the physical realization of quantum information processing (QIP) is the precise control of the system.
In order to achieve this, we must be able to characterize the noise sources affecting experimental setups. A first answer to this problem was given by quantum state and quantum process tomographies (QST, QPT), which allow one to reconstruct the dynamics occurring in the probed system during a given time step \cite{QFTtomo}. But beyond a few qubits, a complete characterization of the errors through this method is not practical, since for a system of $n$ qubits it would require $O(2^{2n})$ measurements \cite{notomo}.

This has motivated the search for protocols showing particular features of the system dynamics but with a scalable implementation,
as in \cite{one_bit_integrability,fid_indicator,Colm,bettelli}. Recently, schemes using random maps have received special attention: just as random numbers play a fundamental role in classical information theory, random unitary operators and random quantum states are also very useful components for quantum information theory and processing \cite{Paz}. Exact random operators are exponentially hard to implement, turning any of these proposals into unscalable ones. 
However, it was recently suggested \cite{pseudorandom} that pseudo-random operators displaying the principal features of random behavior can be constructed efficiently.

Following this, efforts have been made to find signatures of the environment in evolutions generated by random operators. 
Emerson et al. \cite{noise_estimation_TrE} have obtained results linking the fidelity decay with the strength of the noise affecting the system. \\

Here we describe how to extract additional information about the errors. In our approach the noise over a time interval $\tau$ is represented by an error
operator $E=\exp(-i G_{\tau})$, where $G_{\tau}$ is the noise generator. The composition of $G_{\tau}$ gives rise to a variety of noise models. 
These models with their assumptions are stated in Sec. II and further illustrated in Appendix A.
Several results were obtained showing an exponential-like fidelity decay, which we report at the end of Sec. II and further develop in Appendix B.
Moreover, we derived an analytical expression for the initial decay rate, under the assumptions of weak noise and relatively good control.
As we will show, this particular result allows us
to determine the magnitude of the terms in $G_{\tau}$ with different Hamming weights for a given subset of the $n$ qubits.
(The Hamming weight of a product operator -i.e., an operator that is a product of the Pauli matrices $\sigma_x$, $\sigma_y$, $\sigma_z$ or the Identity operator $\Id$
for each qubit- is the number of factors which are different from $\Id$.)
These results are presented in Sec. III, with some more general formulae deferred to Appendix C. 
Sec. IV is devoted to practical implementation issues.
Finally, in Sec. V, we present our conclusions and an outline of future work. 

\section{Description of the noise model and general results}

We start with the standard definition of fidelity after motion reversal for a pure state (also called Loschmidt Echo),
\begin{equation}
f(t)=\abs{\Sandwich{\Psi_0}{U^{\dag}(t) U_P(t)}{\Psi_0}}^2
\label{le}
\end{equation}
where $U$ is the ``perfect'' evolution of the system, while $U_P$ represents a perturbation of $U$.
As it has been shown before (see \cite{fid_indicator,LE} and references therein), the fidelity after motion reversal encodes information about the perfect evolution and/or about the perturbation. Our aim is to obtain information about the perturbation, which for us consists in the errors resulting from uncorrelated imperfect implementation of gates and environmental/external effects. The question is how to choose the perfect evolution in order to only reveal information about the errors. Random operators are a reasonable choice, since randomness enables the exploration of the whole Hilbert space while removing the effect of irrelevant parameters (such as particular choices of initial states). In addition, the use of random operators suppress the effect of possible time-correlations between the terms in the noise generator.

\begin{figure}
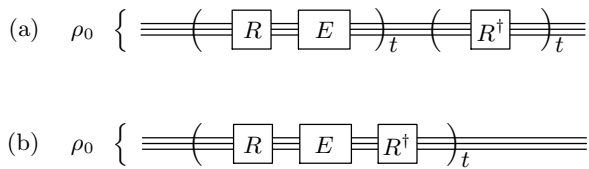

  \begin{center}
    \begin{qcircuit}{21}{7}{1}{25}
      \qbrace{0}{0}{}{(a) \hspace{0.2cm} $\rho_0$}{1}
      \qmultiline{0}{3} 
      \qparenthesis{0}{0}
      \qsingle{0}{$R$} \qskip[0.3] 
      \qsingle{0}{\hspace{.5em}$E$\hspace{.5em}}
      \qendparenthesis{0}{0}{t}
      \qparenthesis{0}{0}
      \qsingle{0}{$R^\dag$} 
      \qendparenthesis{0}{0}{t}
    \end{qcircuit}
  \end{center}
  \begin{center}
    \begin{qcircuit}{21}{7}{1}{25}
      \qbrace{0}{0}{}{(b) \hspace{0.2cm} $\rho_0$}{1}
      \qeraseline{1}
      \qmultiline{0}{3} 
      \qparenthesis{0}{0}
      \qsingle{0}{$R$} \qskip[0.3] 
      \qsingle{0}{\hspace{.5em}$E$\hspace{.5em}} \qskip[0.3] 
      \qsingle{0}{$R^\dag$} 
      \qendparenthesis{0}{0}{t}
    \end{qcircuit}
  \end{center}
\caption{Circuit representation of the algorithm. The operators $R$ are randomly drawn in every step. 
The errors $E$ are also allowed to vary in each step. 
(a) The algorithm seen as a fidelity decay scheme, as given by eqs. (\ref{def_fid}) \& (\ref{def_fid2}). (b) Equivalent algorithm, seen in a twirling
fashion. The equivalence of both designs is due to the average over the random operators $R$ which are taken to be invariant under the Haar measure.}
\label{fig:circuit_fid}
\end{figure}

To incorporate initial states other than pure states, we took a generalization of (\ref{le}) for density matrices
\begin{eqnarray}
f(t) & = & Tr[ \rho_{mr}(t) \rho_0 ] \label{def_fid} \\
\textrm{with }\rho_{mr}(t) & = & R_1^{\dag} \ldots R_t^{\dag}\, E_t\, R_t \ldots E_1\, R_1\, \rho_0 \nonumber\\
                           &   & {} \times R_1^{\dagger}\, E_1^{\dagger} \ldots R_t^{\dagger}\, E_t^{\dagger}\, R_t \ldots R_1 \label{def_fid2}
\end{eqnarray}
where $\rho_0$ is the initial state of the system, the operator $E_t$ represents the errors cumulated during one step of the algorithm at time $t$ and $R_t$ represents some random operator in the Hilbert space $\mathscr{H}_N$ of $n$ qubits ($N=2^n$). 
The subscript $mr$ stands for \emph{motion reversed}. The algorithm is depicted in Fig. \ref{fig:circuit_fid} (a). 
We take the operators $E_t$ to be unitary.
Since we use random operators, we are actually concerned with the average fidelity $\mean{f(t)}$: the brackets denote averaging 
over the considered group of random operators, which we take to be invariant under the Haar measure.
This average makes this fidelity decay scheme equivalent to a twirling scheme \cite{twirling}, as depicted in Fig. \ref{fig:circuit_fid} (b).
This equivalence can be derived from the unitary invariance of the Haar measure. Being a step-by-step twirling of $E$, our proposal breaks up 
the action of $E$ making the state at one step depend 
only on the previous one. Therefore, we can already conjecture an exponential-like decay for $\mean{f(t)}$, as in the work by Emerson et al. \cite{noise_estimation_TrE}.

Defining the fidelity between two mixed states as in (\ref{def_fid}) may seem arbitrary; there is no unique generalization of (\ref{le}) for initial states that are not pure 
(see for example \cite{jozsa}). It will become clear later that this mathematical 
entity reflects the quantity we measure in order to implement our proposal, and for convenience we shall call it fidelity throughout this work.\\

\begin{figure}
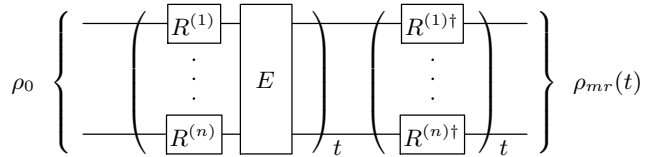

\vspace{0.5cm}
  \begin{center}
    \begin{qcircuit}{21}{7}{3}{25}
      \qbrace{0}{2}{}{$\rho_0$}{1}
      \qeraseline{1}
      \qparenthesis{0}{2}
      \qsingle{0}{$R^{(n)}$} \qsame \qnogate{1} \qsingle{2}{$R^{(1)}$} \qskip[0.3]
      \qmultiple{0}{2}{\hspace{.5em}$E$\hspace{.5em}}
      \qendparenthesis{0}{2}{t}
      \qparenthesis{0}{2}
      \qsingle{0}{$R^{(n)\dagger}$} \qsame \qnogate{1} \qsingle{2}{$R^{(1)\dagger}$}
      \qendparenthesis{0}{2}{t}
      \qendbrace{0}{2}{$\rho_{mr}(t)$}
    \end{qcircuit}
  \end{center}
\caption{Circuit representation of the algorithm, eqs. (\ref{def_fid}) \& (\ref{def_fid2}), choosing individual rotations as the random operators.}
\label{fig:circuit_fid2}
\end{figure}

We chose the random operators to be $R_s = R^{(1)}_s \otimes R^{(2)}_s \otimes \hdots R^{(n)}_s$, where $R^{(j)}_s$ is a random rotation of the qubit $j$ (the resulting algorithm 
is depicted on Fig. \ref{fig:circuit_fid2}). Another possible choice, consisting of $R_s$ being uniformly drawn from $U(N)$, has already been studied in \cite{noise_estimation_TrE}; this led 
to general and closed results for $\mean{f(t)}$, essentially showing a universal exponential decay depending only on the purity of the initial state (that is $\Tr[\rho_0^2]$) and on the global
strength of the noise, quantified by the trace of the superoperator describing the non-unitary dynamics. Although this is a useful analytical result, 
this strong randomization scheme doesn't yield any information
on the noise structure, hinting at the usefulness of a weaker form of randomization.

The $R^{(j)}_s$ are drawn uniformly from $SU(2)$ with respect to the invariant Haar measure. Their expression in the computational basis is \cite{rotation}
\begin{displaymath}
R^{(j)}_s = \left (
\begin{array}{cc}
\cos(\phi_{j,s}) e^{i\psi_{j,s}} & \sin(\phi_{j,s}) e^{i\chi_{j,s}}\\
-\sin(\phi_{j,s}) e^{-i\chi_{j,s}} & \cos(\phi_{j,s}) e^{-i\psi_{j,s}} 
\end{array}
\right )
\end{displaymath}
with $\psi$ and $\chi$ drawn uniformly from the interval $[0,2\pi)$,
and $\phi=\arcsin\left(\sqrt{\xi}\right)$ with $\xi$ uniformly distributed in $[0,1)$.

Notice that these random operators can be efficiently implemented, since they are single-qubit operations with a suitable gate decomposition \cite{nielsen}. 
In this respect, we're not affected by the efficiency issue that arises in the implementation of random operators in $U(N)$ with arbitrarily large $N$, which leads to the use
 of pseudo-random operators.\\ 

We assume we have relatively good control of the system under study, so the random rotations can be implemented with sufficient accuracy and then the errors are only present in the error operators $E_t$.
This is a reasonable hypothesis since we can always make the magnitude of the errors stemming from $E$ relatively larger than the ones in the random rotations by increasing 
the implementation time of $E$. 
A fair exploration of the errors affecting the system can be achieved
by trying to implement the identity operator $\Id$, i.e., trying to prevent any evolution of the system for a certain time $\tau$.

Thus we consider $E$ to be a deviation from $\Id$ in
the form
\begin{equation}
E=\exp\left(-iG_{\tau}\right) = \exp\left(-i\sum_{l=1}^{N^2-1} \chi_l O_l \right)
\label{generator}
\end{equation}
where $G_{\tau}$ is the generator of $E$. We can regard $E$ to be a residual operator resulting from the action of the noise during a time $\tau$ through an effective Hamiltonian $\hbar G_{\tau}/\tau$, for the time step under consideration. Without losing generality, the generator is decomposed in the product operator basis $O_l$:
\begin{equation}
O_l = \bigotimes_{j=1}^n O_l^{(j)}
\label{O_l}
\end{equation}
where each $O_l^{(j)}$ is an operator in the space of qubit $j$ and it is either a Pauli matrix or the Identity (thus the $O_l$ are hermitian), but at least one factor in each $O_l$ is a Pauli matrix (thus the $O_l$ are traceless). For the error operators $E$ to be unitary, the coefficients $\chi_l$ must be real numbers.\\

We can distinguish three classes of effective noise models depending on the time variation of $E$:
\begin{itemize}
\item[C -] Coherent: $E$ remains the same at all times.
\item[IL -] Incoherent with long correlation time: $E$ remains the same during 
approximately the time required to implement one realization of $f(t)$.
\item[IS -] Incoherent with short correlation time: 
$E$ changes from gate to gate.
\end{itemize}
In types IL and IS, the change of $E$ will be given by a change of the coefficients $\chi_l$ in $G_{\tau}$; these are randomly drawn according to a given distribution $P(\{\chi_l\})$.
Coherent noise gives unitary errors. Incoherent scenarios give rise to non-unitary errors; we call them ``incoherent'' since the superoperator arising in these cases has the form
of an (incoherent) average over the parameters characterizing the unitary operation:
\begin{equation}
S_E(\rho) = \int E \left( \{\chi_l\} \right) \rho E^{\dagger}(\{\chi_l\}) \ P(\{\chi_l\}) \ \ud \chi_l
\end{equation}
Notice that in each realization, although at each step the action of the error operator is unitary, its variation in time introduces a net non-unitary operation 
when averaging different realizations to obtain $\langle f(t) \rangle$.
A brief illustration of these models is given in Appendix A for the one qubit case.\\

We expect that some of all the terms we allow in the noise generator will be negligible.
We will analyze $G_{\tau}$ for truncated sums over multi-body terms $O_l$ up to a given Hamming weight. 
For this we introduce a more specific labeling of the terms in $G_{\tau}$. When necessary, we shall denote the $\chi_l$ as $\chi_{j,k \ldots}^{p,q \ldots}$,
where $j,\ k, \ldots$ label qubits, and together with $p,\ q, \ldots = \{ x,\ y,\ z \}$, they indicate that that particular term is a product of the Pauli matrices $\sigma_p$ for 
qubit $j$, $\sigma_q$ for qubit $k$, etc., and the rest of the factors are just the Identity for the other qubits. Therefore, the one-body terms (Hamming weight $1$) go with coefficients
$\chi_j^x$, $\chi_j^y$, $\chi_j^z$, two-body terms (Hamming weight $2$) are $\chi_{j,k}^{p,q}$, etc. To avoid double counting of multi-body terms, the labeling of the qubits must obey
$j < k < \ldots$ and so on.\\

As we already mentioned, the non-negligible coefficients can in general be drawn from any
given distribution $P(\{\chi_l\})$. We studied $\langle f(t) \rangle$ for some specific cases, including only one-body terms (analytical and numerical results) and one-body and two-body 
terms  (numerically). The two distributions considered were: 1) a constant distribution ($\chi^p_j=\alpha$, $\chi_{j,k}^{p,q}=\beta$  $\forall\ j,\ p,\ k,\ q$);
2) each $\chi^p_j$ (resp. $\chi_{j,k}^{p,q}$) randomly drawn from a Gaussian distribution with mean value $\alpha$ ($\beta$) and standard deviation 
$\sigma_{\alpha}$ ($\sigma_{\beta}$).
We will refer to the coefficients $\chi_l$ or to $\alpha$, $\beta$, $\sigma_{\alpha}$, $\sigma_{\beta}$ collectively as 
the ``noise strength'' $\chi$, and the powers of $\chi$ will include any monomial combination of degree equal to the given power.

\begin{figure}[h]
\includegraphics[width=0.38\textwidth, angle=270]{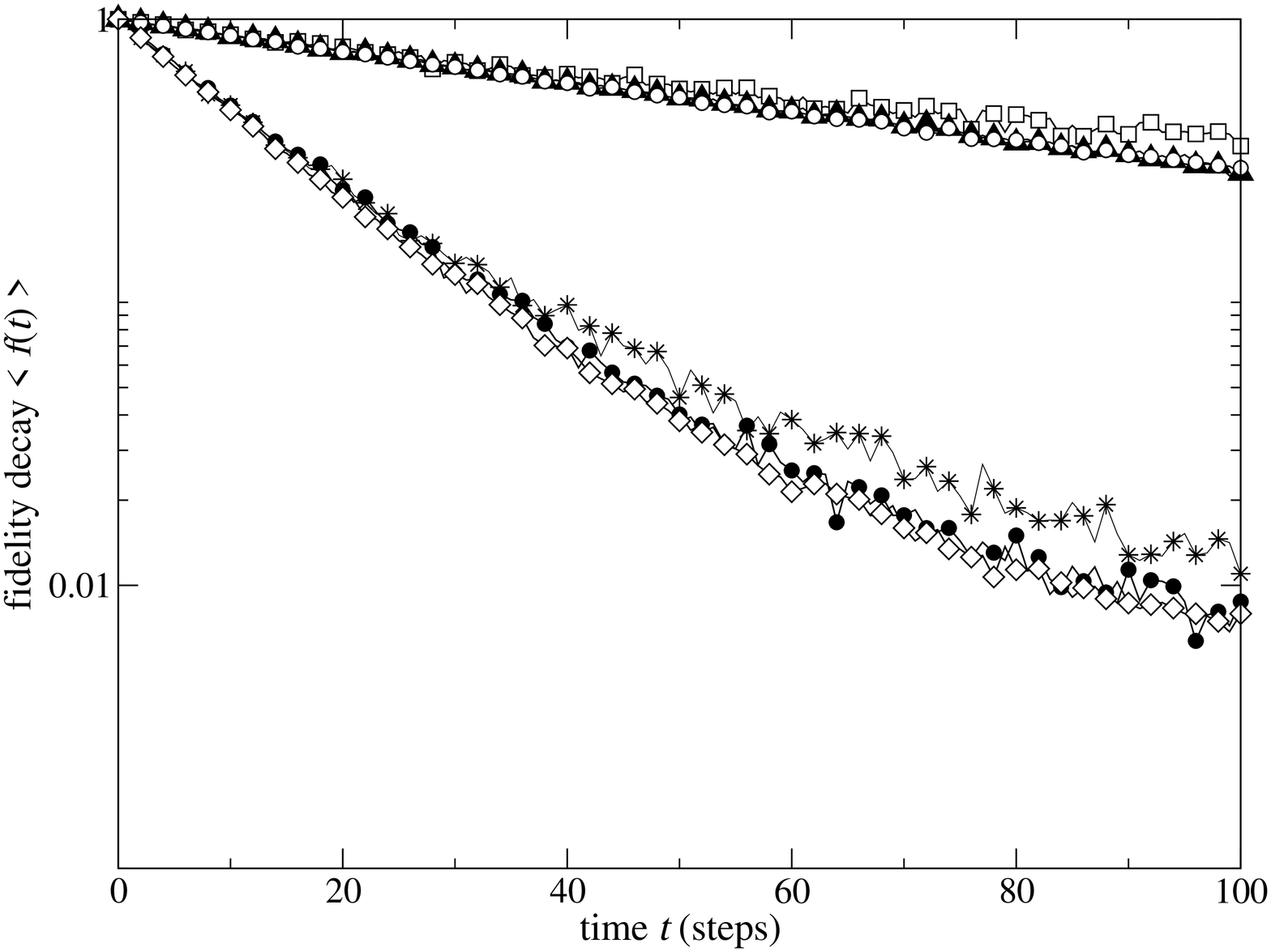}
\caption{\small \label{fig:curves005} Examples of the numerical calculation of $\langle f(t) \rangle$ with 100 realizations for 8 qubits, 
initially all in the $| 0 \rangle$ state. Here we 
took $G_{\tau}$ with one-body terms only ($\square, \blacktriangle, \Circle$) and with one-body and all the two-body terms (*, $\bullet, \diamondsuit$).}
\begin{flushleft}
\vspace{-0.3cm}
\small
$\square$: type IL, $P$ Gaussian with  $\alpha=\beta=0,\ \sigma_{\alpha}=0.05, \sigma_{\beta}=0$.\\
$\blacktriangle$: type C, $P$ constant with $\alpha=0.05,\ \beta=0$. \\
$\Circle$: type IS, $P$ Gaussian with $\alpha=\beta=0,\ \sigma_{\alpha}=0.05, \sigma_{\beta}=0$. \\
*: type IL, $P$ Gaussian with $\alpha=\beta=0,\ \sigma_{\alpha}=\sigma_{\beta}=0.05$. \\
$\bullet$: type C, $P$ constant with $\alpha=\beta=0.05$. \\
$\diamondsuit$: type IS, $P$ Gaussian with $\alpha=\beta=0,\ \sigma_{\alpha}=\sigma_{\beta}=0.05$.
\end{flushleft}
\vspace{-0.3cm}
\begin{tabular*}{8cm}{c}
\hline
\end{tabular*}
\end{figure}

Numerical calculations of $\langle f(t) \rangle$ show an exponential-like decay. In particular we observe: 
\begin{itemize}
\item[$i$)] Linear initial decay: $\langle f(t) \rangle \approx f_0 (1 - \gamma t)$ for $t$ sufficiently small and with $f_0=\Tr[\rho_0^2]$ 
(we shall call $\gamma$ the initial decay rate); 
\item[$ii$)] Constant long-time limit: $\langle f(t) \rangle \rightarrow 1/N$ for $t \rightarrow \infty$.
\end{itemize}
The scales ``$t$ sufficiently small'' or $t \rightarrow \infty$ are set by the strength of the noise $\chi$.
Our numerical calculations ranged up to a strength of $0.4$. For 
higher strength, the saturation value $1/N$ is reached in only a few steps and not much can be extracted from this fidelity decay. 
An example to illustrate these calculations is given in Fig. \ref{fig:curves005}.\\

Analytical expressions of $\langle f(t) \rangle$ can be obtained if we consider one-body terms only and a separable initial state.
Our closed results are in exact agreement with the behavior described above. The slope of the decay is of order $O(\chi^2)$ and is
the same for the different types of noise C, IL and IS.
The analytical results were derived with mathematical tools developed in \cite{Samuel}, a good presentation thereof can be found in \cite{Brouwer}. 
This approach has been used in \cite{Frahm} to study the fidelity decay of perturbed quantum chaotic maps.

We defer further details about the calculations for the full $\langle f(t) \rangle$, since the most interesting results arise from our studies of
the initial decay rate of $\langle f(t) \rangle$. The reader is referred to Appendix B for a more complete report on the former.\\

To conclude this Section we revisit the following point: our work indicated that the fidelity decay after motion reversal is initially linear in time,
also in agreement with the results already published by Emerson et al. \cite{noise_estimation_TrE}. 
While this seems to contradict previous well established results reporting a universal quadratic decay \cite{LE}, these two statements do not in fact contradict each other, since the 
random dynamics studied here is not considered by previous work, which presumes the use of a constant evolution operator. 
In our case, the evolution is given by random rotations which vary in each step and in each realization, and we then study the evolution of an
ensemble-averaged state. Even when some relations between our error operator $E$ and the perturbation can be drawn (which originally motivated the use of 
a fidelity decay scheme), the nature of the calculation is different.\\

\section{The initial decay rate $\gamma$}

\begin{figure}[h]
\centering
\includegraphics[width=0.38\textwidth,angle=270]{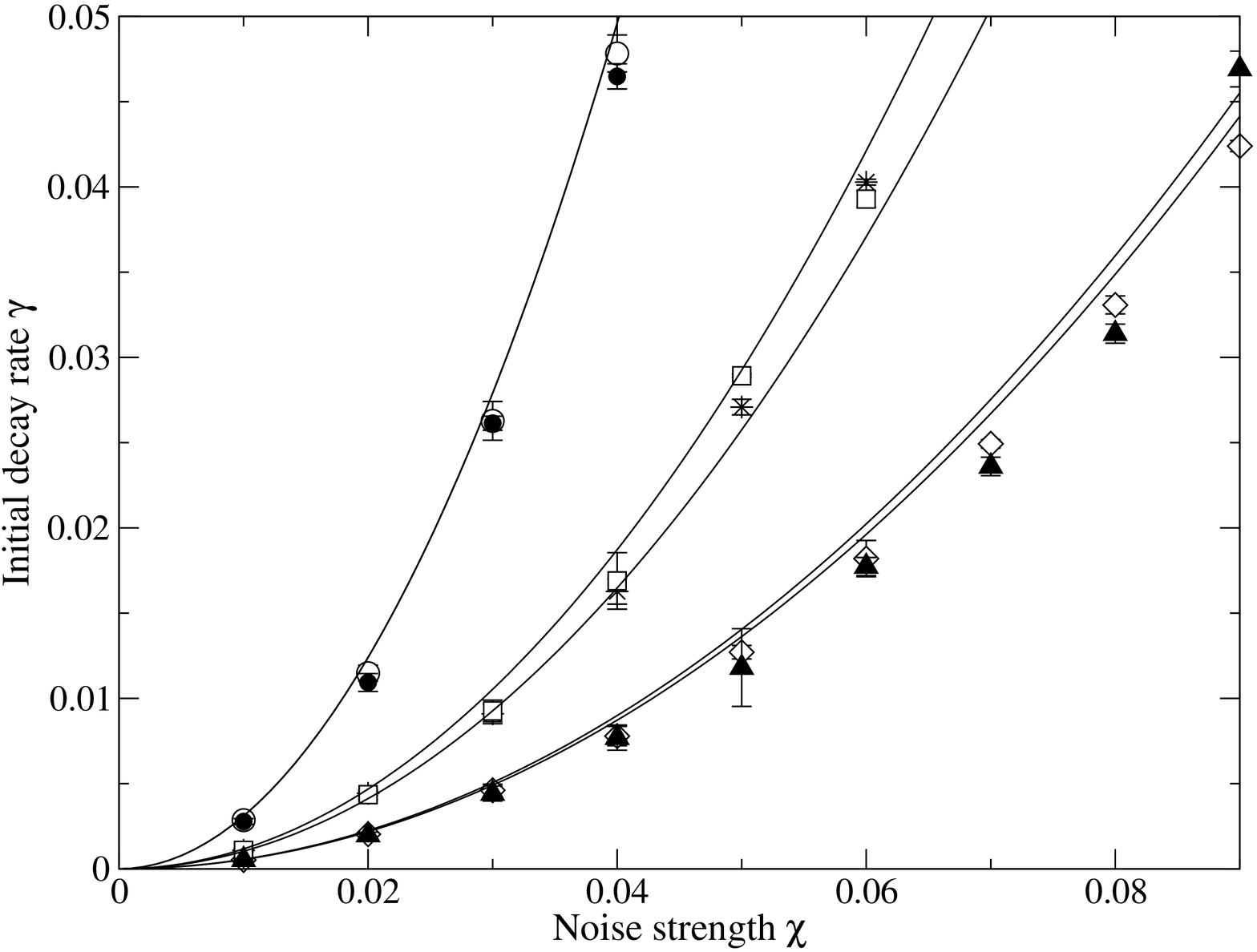}
\caption{\label{fig:gammas} Values of the initial decay rate $\gamma$ obtained with a linear fitting of $\mean{f(t)}$ in the small $t$ regime. In practice, this linear regime was given
by all the points $t$ with $\mean{f(t)} > f_{lim}$. Here $f_{lim}=0.9$ and $n=8$. Solid lines: the quadratic fitting.}
\begin{flushleft}
\vspace{-0.3cm}
\small
$\Circle$: type C, $P$ constant with $\alpha=\beta=\chi$; all the terms in $G_{\tau}$. \\
$\bullet$: type IL, $P$ Gaussian with $\alpha=\beta=0$, $\sigma_{\alpha}=\sigma_{\beta}=\chi$; all the terms in $G_{\tau}$. \\
*: idem $\Circle$, but only one-body and first-neighbor terms. \\
$\square$: idem $\bullet$, but only one-body and first-neighbor terms. \\
$\diamondsuit$: idem $\Circle$, but only one-body terms. \\
$\blacktriangle$: idem $\bullet$, but only one-body terms.
\end{flushleft}
\vspace{-0.3cm}
\begin{tabular*}{8cm}{c}
\hline
\end{tabular*}
\end{figure}

Numerical and analytical evidence supports the conjecture that the initial decay is linear in $t$, with an initial decay rate $\gamma$. 
Moreover, this initial decay rate depends only on the noise strength (the magnitude of the $\chi_l$) and not on the particular time variation of $E$.
This can be seen in Fig. \ref{fig:curves005}, where the initial decay is the same for the different types 
C, IL and IS (described in Sec. II) as long as the general noise strength is the same.
In addition, the dependence on the strength is quadratic (Fig. \ref{fig:gammas}).

The first clear evidence comes from the initial slope of the analytical expressions we obtained;
this result is of course limited to noise with only one-body terms. For noise including two-body terms, we fitted the initial decays (as shown in
Fig. \ref{fig:gammas}), obtaining a quadratic dependence on the governing parameter. 

We see then that the initial decay rate is a measure of the noise strength independently of the type of noise.
By defining precisely the initial decay rate $\gamma$ in terms of the fidelity after the first step,
\begin{equation}
\gamma = 1 - { \langle f(t=1) \rangle \over f_0 }
\end{equation}
we can actually obtain an analytical expression for the initial decay rate
up to second order in $\chi$, for a separable initial state. We can also prove that the third order in $\chi$ vanishes. Notice that for this calculation 
we went back to a general noise model with multi-body terms in $G_{\tau}$ 
($\chi_l \neq 0\ \forall \ l$ in principle). For an initial state where each qubit is in a pure state ($f_0 = 1$), we get
\begin{eqnarray}
\gamma &=& c_1\sum_{j=1}^{n} (\chi^*_{j})^2 + c_2 \sum_{k > j = 1}^n (\chi^*_{j,k})^2  \nonumber \\
&+& c_3 \sum_{g > k > j = 1}^n (\chi^*_{j,k,g})^2 + \ldots + O(\chi^4)
\label{allpure}
\end{eqnarray}
with $c_{\nu} = 1 - 1/3^{\nu} $. In (\ref{allpure}) we have defined the collective coefficients
\begin{eqnarray}
(\chi^*_j)^2 = \sum_{p=x,y,z} (\chi_j^p)^2; \ (\chi^*_{j,k})^2 = \sum_{p,q=x,y,z} (\chi_{j,k}^{p,q})^2; \ \textrm{etc.} 
\label{collective}
\end{eqnarray}
and $\nu=1, 2, 3, \ldots, n$ for a collective coefficient corresponding to terms with Hamming weight $\nu$.

In the case where the coefficients are fluctuating over time the relevant quantity is the average 
\begin{equation}
\langle \gamma \rangle_P = \int \gamma(\chi) P(\chi) \ud \chi
\label{gamma_P}
\end{equation}
since when we average the realizations of the random rotations, we also average realizations of the fluctuating coefficients.
For example for the Gaussian distribution we described before we would have $\int (\chi^*_j)^2 P(\chi) \ud \chi = 3 (\alpha^2 + \sigma_{\alpha}^2)$, etc.
Therefore for any distribution $P$ the same equations hold, with the collective coefficients properly replaced by the strength parameters characterizing $P$.\\

\begin{figure}
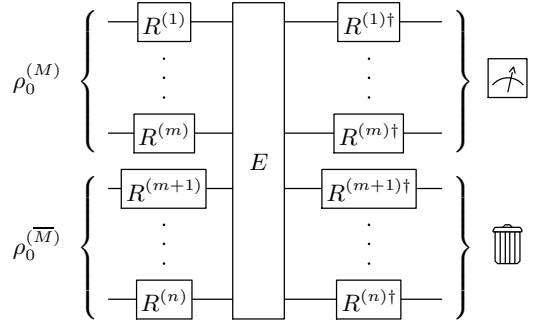

\vspace{0.5cm}
  \begin{center}
    \begin{qcircuit}{21}{5}{6}{25}
      \qbrace{0}{2}{}{$\rho_0^{(\overline{M})}$}{1}
      \qbrace{3}{5}{}{$\rho_0^{(M)}$}{1}
      \qeraseline{1}
      \qeraseline{4}
      \qsingle{0}{$R^{(n)}$} \qsame \qnogate{1} \qsingle{2}{$R^{(m+1)}$} \qsame 
      \qsingle{3}{$R^{(m)}$} \qsame \qnogate{4} \qsingle{5}{$R^{(1)}$} \qskip[0.7]
      \qmultiple{0}{5}{\hspace{.5em}$E$\hspace{.5em}} \qskip[1]
      \qsingle{0}{$R^{(n)\dagger}$} \qsame \qnogate{1} \qsingle{2}{$R^{(m+1)\dagger}$} \qsame
      \qsingle{3}{$R^{(m)\dagger}$} \qsame \qnogate{4} \qsingle{5}{$R^{(1)\dagger}$} 
      \qendbrace{0}{2} {\hspace{0.1cm}}
      \qendbrace{3}{5} {\hspace{0.1cm}} 
      \qskip[1.5] \qmeasure{4} \qsame \qtrashcan{1}
    \end{qcircuit}
  \end{center}
\caption{Circuit representation of the algorithm in its final, experimental version.}
\label{fig:circuit_fid_partial}
\end{figure}

Eq. (\ref{allpure}) shows a decay rate $\gamma$ that is a weighted sum of the collective coefficients of $G_{\tau}$. However, we are rather
interested in obtaining a characterization distinguishing these coefficients. If we chose other initial states, the weights of
the collective coefficients change; moreover, some vanish if some qubits are initially in the maximally mixed state $\Id /2$. To make use of this feature efficiently, 
we can calculate the fidelity of the state of just a few qubits. Let's call $M$ the set of $m$ qubits that is going to be measured ($m\le n$),
and $\overline{M}$ its complementary (Fig. \ref{fig:circuit_fid_partial}). Thus we have
\begin{eqnarray}
\langle f^{(M)}(t) \rangle &=& Tr_M[ \rho^{(M)}_{mr}(t) \rho^{(M)}_0 ]
\label{def_partial_fid}
\end{eqnarray}
where we denote the reduced density matrices by $\rho^{(X)}=\Tr_{\overline{X}}[\rho]$. 
Correspondingly, we denote as $\gamma^{(M)}$ the initial decay rate of $\mean{f^{(M)}(t)}$.

It can be shown that $\gamma^{(M)}$ is independent of the initial state of the qubits not being measured. This is indeed a desirable feature
since we then don't have to worry about experimentally initializing them -as long as the separability of the initial state of the $m$ qubits in
$\mathscr{H}^{(M)}$ is guaranteed. We show below the results for measuring the coefficients of arbitrary sets of one, two and three qubits, 
which we have labeled $a$, $b$ and $c$;
these qubits are initially in an arbitrary pure state. More general formulae is given in Appendix C; the following will suffice to set the basis
for our proposal.
\begin{eqnarray}
\rho_0^{(a)} &=& |\varphi_a\rangle \langle \varphi_a| \nonumber \\
\gamma^{(a)} &=& {2 \over 3} \left( \rule{0cm}{0.5cm} \right. 
(\chi^*_a)^2 + \sum_{j \neq a} (\chi^*_{a,j})^2 + \mathop{\sum_{k > j}}_{j, k \neq a} (\chi^*_{a,j,k})^2 + \ldots \left. \rule{0cm}{0.5cm} \right) \nonumber \\
&+& O(\chi^4)
\label{gamma_a}
\end{eqnarray}
\begin{eqnarray}
& &\rho_0^{(a)} \otimes \rho_0^{(b)} = |\varphi_a\rangle \langle \varphi_a| \otimes |\varphi_b\rangle \langle \varphi_b|  \nonumber \\
& & \gamma^{(a,b)} = \frac{2}{3} \left( \rule{0cm}{0.7cm} \right. (\chi^*_a)^2 + (\chi^*_b)^2 + \sum_{j \neq a, b} \left[ (\chi^*_{a,j})^2 + (\chi^*_{b,j})^2 \right]  + \nonumber \\
& & \ \ + \mathop{\sum_{k > j}}_{j, k \neq a, b} \left[ (\chi^*_{a,j,k})^2 + (\chi^*_{b,j,k})^2 \right] + \ldots \left. \rule{0cm}{0.7cm} \right)  \nonumber \\
& & \ \ +\ \ {8 \over 9} \left( (\chi^*_{a,b})^2 + \sum_{j \neq a, b} (\chi^*_{a,b,j})^2 + \ldots \right) + O(\chi^4)
\label{gamma_ab}
\end{eqnarray}
\begin{eqnarray}
& & \rho_0^{(a)} \otimes \rho_0^{(b)} \otimes \rho_0^{(c)} = |\varphi_a\rangle \langle \varphi_a| \otimes |\varphi_b\rangle \langle \varphi_b| \otimes |\varphi_c\rangle \langle \varphi_c| \nonumber \\
& & \gamma^{(a,b,c)} = {2 \over 3} \left( \rule{0cm}{0.7cm} \right. \rule{0cm}{0.5cm}(\chi^*_a)^2 + (\chi^*_b)^2 + (\chi^*_c)^2 + \nonumber \\
& & \ \ \ +  \sum_{j \neq a, b, c} \left[ (\chi^*_{a,j})^2 + (\chi^*_{b,j})^2 + (\chi^*_{c,j})^2  \right]+ \nonumber \\
& & \ \ \ + \mathop{\sum_{k > j}}_{j,k \neq a, b, c} (\chi^*_{a,j,k})^2 + (\chi^*_{b,j,k})^2 + (\chi^*_{c,j,k})^2 + \ldots \left. \rule{0cm}{0.7cm} \right) \nonumber \\
& & \ \ \ + \ \ {8 \over 9} \left( \rule{0cm}{0.7cm} \right. (\chi^*_{a,b})^2 + (\chi^*_{a,c})^2 + (\chi^*_{b,c})^2 \nonumber \\
& & \ \ \ + \sum_{j \neq a, b, c} (\chi^*_{a,b,j})^2 + (\chi^*_{a,c,j})^2 + (\chi^*_{b,c,j})^2 + \ldots \left. \rule{0cm}{0.7cm} \right) \nonumber \\
& & \ \ \ +\ \ {26 \over 27} \ (\chi^*_{a,b,c})^2 + O(\chi^4)
\label{gamma_abc}
\end{eqnarray}
It's expected that for small errors (or equivalently, for $\tau$ small enough) only terms with low Hamming weight will be present in $G_{\tau}$. This being the case,
by measuring the initial decay rate of a few qubits, the value of selected coefficients of $G_{\tau}$ can be extracted. For example, if terms with Hamming weight $\geq 3$ are negligible,
the combination
\begin{equation}
\gamma^{(a)} + \gamma^{(b)} - \gamma^{(a,b)}  = {4 \over 9} (\chi^*_{a,b})^2
\label{ab_coefficient}
\end{equation}
allows us to establish whether any two-body term between an arbitrary pair of qubits $a$  and $b$ is present in $G_{\tau}$.
Notice that the measurements return the value of a given coefficient averaged over the distribution $P(\chi^p_j,\chi_{j,k}^{p,q})$ (refer to eq. (\ref{gamma_P})), 
giving thus its strength according to the parameters of $P$.
In this way we can probe any two-qubit collective coefficient we are interested in, or conduct a fair sampling of some of them.

The systematic protocol to measure the collective coefficients for one and two-body terms is the following:
\begin{itemize}
\item[1.] Measure the $n$ initial decay rates $\gamma^{(j)}$ for all the qubits individually. That is: apply one step of the algorithm 
given in Fig. \ref{fig:circuit_fid_partial} measuring only qubit $j$ initially in a pure state, thus obtaining \mbox{$\langle f^{(j)}(t=1) \rangle$}.
From this the initial decay rate can be obtained by subtracting $f_0^{(j)}=1$. 
\item[2.] Measure the $n(n-1)/2$ initial decay rates for all the possible pairs, $\gamma^{(j,k)}$. This is just as explained above
but now measuring qubits $j$ and $k$, each initially in a pure state.
\item[3.] With this data, all the two-body coefficients can be determined using (\ref{ab_coefficient}).
\item[4.] All the one-body coefficients can be extracted by subtracting the two-body coefficients from the initial decay rates
of one qubit, according to (\ref{gamma_a}).
\end{itemize}
This implementation does not distinguish between different product operators $O_l$ for a given subset of the $n$ qubits (i.e., between $X$, $Y$, and $Z$ directions), 
since all the corresponding coefficients add up to form the collective coefficients as expressed in (\ref{collective}).\\

If terms with higher Hamming weight are present in the generator, we can extend the method, but of course the number of initial decay rates required to map out the $\chi_l$ 
increases, eventually becoming exponential in $n$. The advantage of this approach is that, when higher order terms are negligible, it makes good use of this fact. 
In addition, it provides a procedure to measure selected coefficients instead of going necessarily for the whole set.
It is possible, for example, to probe the importance of three-body terms in $G_{\tau}$ (neglecting
terms with Hamming weight $\geq 4$) with the combination
\begin{equation}
\gamma^{(a)} + \gamma^{(b)} + \gamma^{(c)} - \gamma^{(a,b)} - \gamma^{(a,c)}- \gamma^{(b,c)}+ \gamma^{(a,b,c)} = {8 \over 27} (\chi^*_{a,b,c})^2
\nonumber
\end{equation}
More details on the analytical calculation of the initial decay $\gamma^{(M)}$ is given in Appendix C. For this we employed the tools already mentioned at the end of Sec. II.

\section{Implementation issues}

Consider the protocol described in Sec. III running on a canonical quantum computer: when we measure the state of the quantum register at the end of the computation, 
we get only a binary number as a result. 
If the initial state is pure so $\rho_0^{(M)}=\ketbra{\varphi}{\varphi}$, and $|\varphi\rangle = \bigotimes_{j \epsilon M}|\varphi_j\rangle$ 
with $|\varphi_j\rangle = |0\rangle \textrm{ or } |1\rangle$, the fidelity can be evaluated in a simple way
by measuring the final state in the computational basis.

The average can be implemented as follows:
let $X$ be the random variable which is equal to $1$ if the result of the measurement of $\rho^{(M)}_{mr}(t)$ gives $\rho_0^{(M)}$ and $0$ otherwise.
The random variables $X_u$ (each respresenting the outcome of the \nth{u} algorithm realization) all have the same distribution 
$\Prob{X_u=1}=\mathrm{E}(X_u)=\mean{f^{(M)}}$, independently of the particular realization.
The mean 
\begin{equation}
\overline{X}=\frac{1}{N_R} \sum_{u=1}^{N_R}X_u 
\end{equation}
(where $N_R$ is the total number of realizations) thus gives an estimation of the average fidelity.
From the Chernoff inequality we can get the minimum number of times $N_R$ we must run the algorithm to achieve a precision $\delta$ on the measure
of the fidelity with an error probability less than $\varepsilon$:
\begin{eqnarray}
& \displaystyle{\Prob{\left|\overline{X}-\mean{f}\right| \ge \delta} \le 2e^{-2 N_R\delta^2} \le \varepsilon} \label{chernoff1} \\
& \displaystyle{\Longrightarrow N_R \ge -\frac{1}{2\delta^2}\ln\frac{\varepsilon}{2}} \label{chernoff2}
\end{eqnarray}
In the scenario where three-body and higher order terms are neglected, we see from equation (\ref{ab_coefficient}) that to be able to distinguish a two-body coefficient 
of magnitude $\beta$ from $0$,
the number of experiments we need to run is of the order of $\ln\varepsilon/\beta^4$,
which doesn't depend on the number of qubits involved and ensures the method's scalability.

Notice again that the average has a double function. It's not only the average over the random rotations but also, when the noise is incoherent, the average of the fluctuating 
coefficients $\chi_l$. 

Eqs. (\ref{chernoff1}) and (\ref{chernoff2}) also hold in the case where $X$ is a continuous variable in the interval $[0,1]$, in this case known also as
the Hoeffding inequality \cite{hoeffding}. This should be noticed when trying to implement this protocol with ensemble measurements instead of projective measurements,
as in NMR QIP.

\section{Conclusions}

As described in \cite{fault-tolerant}, fault-tolerant quantum computing requires the magnitude of the noise affecting the implementation of a gate to be smaller than a certain critical value. 
The quantity measuring the noise magnitude and its threshold value depend on the structure of the noise, where by ``structure'' we mean which multi-body terms are negligible
and how this scales with the number of qubits. 
Fault-tolerant thresholds of this type are formulated in terms of a Hamilitonian $\mathcal{H}$ responsible for the errors in the computation, acting for the time $t_0$ required
to implement a gate. The generator $\mathcal{H} \hspace{0.05cm} t_0 / \hbar$ includes the interaction with an external environment (it thus generates both unitary and non-unitary errors), and has support on the system's space $\mathscr{H}_N$ as well as outside of it (the environment's space). In practice, however, we expect to have access only to the system's space; the intention of our approach is actually to characterize a generator $\mathcal{G}_N$ resulting from the action of $\mathcal{H}$ in the system of $n$ qubits.\\

We have presented here a protocol to analyze the noise structure with at most two-body terms in the generator. The method can be extended if higher order multi-body terms are present,
at the price of compromising its scalability. Notice that in any case the method offers a way to probe the importance of these higher order terms.

We believe our choice of 
the identity operator as primordial gate will give a fair idea of the terms present in the generator, keeping in mind that implementing $\Id$ means implementing a time-suspension 
sequence \cite{time_suspension} that in principle is composed of several gates modulating the internal Hamiltonian of the system.
If we want to analyze the structure of the noise resulting from the implementation of a particular gate $U_g$, we can easily account for this
by implementing $E$ as $E=U_g U_g^{\dagger}$ - its implementation will be as perfect as our ability to reverse it. \\


The main advantage of our proposal is its scalability, but another rather important feature is that its outcome gives directly 
the generator $G_{\tau}$ of the errors over a time $\tau$, close to
the $\mathcal{H}$ referred to in the fault-tolerance analysis. 
This is one step ahead of the QPT approach, whose outcome is essentially a superoperator, from which $\mathcal{H}$ still has to be extracted. 
At this point we should mention that this direct link between measurable quantities and the terms in an effective generator $\mathcal{G}_N$ resulting from the interaction
with an environment is well known in the theory of relaxation in NMR systems, where the $1/T_2$ initial decay rates encode the magnitude of 
these terms and the correlations between them \cite{Ernst}.\\

However, as we have mentioned before, our proposal relies on certain assumptions which limit its reach. One assumption is that 
terms with high Hamming weight can be neglected; this is nevertheless a reasonable one. 
Our method relies on this fact in order to achieve scalability, differing from QPT in that it can make good use of this assumption.
Another assumption is that we have taken the error operators $E$ to be unitary, thus confining the non-unitary errors we considered to a subset of unital processes. 
This assumption is again reasonable
considering the time scale of typical non-unital processes (the so-called $T_1$-processes or relaxation); unital processes are expected to occur faster thus the corresponding coefficients are 
expected to be larger.

Our current work is directed to generalize our proposal to include more general noise scenarios; also, further studies should concentrate in accessing the information we need from $\mathcal{H}$ (the full set of interaction terms for system plus environment) when only the system's space is available to us.\\

This work was supported in part by the National Security Agency (NSA) under Army Research Office (ARO) contract number W911NF-05-1-0469,
by the Air Force Office of Scientific Research and
by a Lavoisier Fellowship from the French Ministry of Foreign Affairs.

\section{Appendix A: Simple model of incoherent noise}

Here we present an illustrative interpretation of our noise scenarios. The case where $E$ is constant (coherent noise) 
evidently reflects the unitary errors that arise due to the implementation of imperfect gates. 

The incoherent cases we have introduced deserve some more analysis. A simple picture of these processes can be obtained by studying the effect of $E$ for only one
qubit, a process that is analytically tractable, being $E=\exp(-i \chi_1 \sigma_z)$.
In order to reflect the incoherent nature of the process, we of course study the evolution of the state averaged over different realizations of $\chi_1$.
Also, to distinguish between incoherent with long (IL) and short (IS) correlation times, we must 
observe the system at a time $t$ for which the IL noise
would remain constant, but nevertheless the IS noise will vary in each step. For these two processes, we have
\begin{eqnarray}
\rho_{IL}(t) &=& \int_{-\infty}^\infty (E(\chi_1))^t \ \rho(0) \ (E^\dag(\chi_1))^t \ P(\chi_1) \ \ud \chi_1\\
\rho_{IS}(t) &=& \int_{-\infty}^\infty \dots \int_{-\infty}^\infty E(\chi_{1,t}) \ldots E(\chi_{1,1}) \rho(0)   E^\dag(\chi_{1,1})  \times \nonumber \\
  &\ldots& E^\dag(\chi_{1,t})  P(\chi_{1,1}) \ldots P(\chi_{1,t}) \ \ud \chi_{1,1} \ldots \ud \chi_{1,t}
\end{eqnarray}
where $P$ is, as mentioned before, a Gaussian distribution centered in $\alpha$ with a deviation $\sigma_{\alpha}$. 

If we start with a general initial state $\rho(0)=(\Id+ \sigma_x \varrho_x + \sigma_y \varrho_y + \sigma_z \varrho_z )/2$, we obtain
\begin{eqnarray}
\rho(t) &=& {\Id \over 2} +  {\sigma_z \over 2} \varrho_z + \left[ {\sigma_x \over 2} (\varrho_x \cos(2 \alpha t) - \varrho_y \sin(2 \alpha t)) \nonumber \right.\\
            &+& \left. {\sigma_y \over 2} (\varrho_x \sin(2 \alpha t) + \varrho_y \cos(2 \alpha t)) \right] \exp(-\delta)
\end{eqnarray}
where $\delta=2 \sigma_{\alpha}^2 t^2$ for the IL case, and $\delta=2 \sigma_{\alpha}^2 t$ for the IS one. In both cases we obtain the same physical process: an
exponential decrease of the transversal polarization (together with a rotation), leaving the longitudinal polarization unchanged. The decay is slower with IS errors, since
their rapid change make them more random and overall less harmful. This non-unitary process has the following Kraus representation
\begin{equation}
\rho(t)=M_1 \rho(0) M_1^\dag + M_2 \rho(0) M_2^\dag
\end{equation}
\begin{eqnarray}
M_1=\left( {1+e^{-\delta} \over 2} \right)^{1/2} \Id \ e^{-i \alpha \sigma_z} \\
M_2=\left( {1-e^{-\delta} \over 2} \right)^{1/2} \sigma_z \ e^{-i \alpha \sigma_z}
\end{eqnarray}
This clearly shows that the process is composed of two parts: on one hand, a rotation (unitary operation) around the $\hat{z}$ axis, related to the fact that the values of $\chi_1$ are centered around $\alpha$. On the other hand, a phase-flip channel with probability $(1-e^{-\delta} )/2 \leq 0.5$ of flipping the qubits.

This kind of process is commonly encountered in liquid NMR QIP, where this simple model reflects the essentials of the errors arising from spurious inhomogeneities in the magnetic field.
In the stochastic limit, the phase-flip is expected to happen with a constant probability. 

\section{Appendix B: The fidelity decay}

\subsection{Analytical results}

Analytical expressions of $\langle f(t) \rangle$ can be obtained if we consider $G_{\tau}$ with one-body terms only,
\begin{equation}
E_t = \bigotimes_{j=1}^n E^{(j)}_t, \ \ E^{(j)}_t = \exp \left(-i \chi_j^* \sigma^{(j)}_{n_j} \right)
\end{equation}
(the actual directions $\hat{n}_j$ are irrelevant) and a separable initial state 
\begin{equation}
\rho_0 = \rho_0^{(1)} \otimes \rho_0^{(2)} \otimes \ldots \otimes \rho_0^{(n)} 
\end{equation}
Under these conditions the fidelity of the whole system is just a multiplication of the fidelities for each qubit:
\begin{equation}
\langle f(t) \rangle =\prod_{j=1}^{n} \langle f^{(j)}(t) \rangle
\label{prod_of_fid}
\end{equation}
We observe that
\begin{eqnarray*}
\lefteqn{ \langle f^{(j)}(t) \rangle = \langle f^{(j)}(R^{(j)}_t, \ldots, R^{(j)}_1) \rangle } \\
&=& \int \ldots \int f^{(j)}(R^{(j)}_t, \ldots, R^{(j)}_1) \ud R^{(j)}_t \ldots \ud R^{(j)}_1
\end{eqnarray*}
where each integral is an average over the normalized Haar measure on $U(2)$. $f^{(j)}(\ldots, R^{(j)}_s, \ldots)$ is a polynomial function of $R^{(j)}_s$ (and $R^{(j)\dagger}_s$ 
of course). 
A method for computing this kind of averages in $U(N)$ is presented in \cite{Samuel, Brouwer}; here we will limit ourselves to state the following particular results
\begin{eqnarray}
& \langle \Tr[ A R B R^{\dagger}] \rangle = {1 \over 2} \Tr[A] \Tr[B] \label{ben1} \\
& \langle \Tr[ \rho R^{\dagger} A R \rho R^{\dagger} B R] \rangle = {1 \over 3} \Tr[A B] \left( 1 - { \Tr[\rho^2] \over 2} \right) \nonumber \\
& + {1 \over 3} \Tr[A] \Tr[B] \left( \Tr[\rho^2] - {1 \over 2} \right) 
\label{ben2}
\end{eqnarray}
where all the operators belong to $\mathscr{H}_2$ and we have used $\Tr[\rho]=1$. Applying this 
formulae we have
\begin{eqnarray}
\langle f^{(j)}(t) \rangle &=& \langle f^{(j)}(t-1) \rangle { | \Tr_j[E^{(j)}_t]|^2 - 1 \over 3} \nonumber \\
&+& \frac{2}{3} - \frac{| \Tr_j[E^{(j)}_t]|^2}{6}
\label{f_step}
\end{eqnarray}
This shows that $f(t)$ only depends on the fidelity at a previous time $t-1$, thus giving an intrinsic exponential-type decay. At the same time,
this also shows that the precise decay law won't be a simple exponential. Even for coherent noise, where $\Tr_j[E^{(j)}_t]$ is the same at all times, 
eq. (\ref{prod_of_fid}) already indicates that we'll have a product of exponentials.

From eq. (\ref{f_step}) it's possible to compute a closed expression for $\langle f(t) \rangle$ in several cases, accounting also
for the time variation of the coefficients in $E_t$ given by $P$.
For coherent errors with a constant distribution $P(\chi_j^*)=\delta(\chi_j^* - \alpha)$ we have
\begin{eqnarray}
\langle f^{(j)}(t) \rangle = {1 \over 2} + \left(f^{(j)}_0 - {1 \over 2} \right) \exp(-\lambda t)
\label{f^j_coherent}
\end{eqnarray}
where
\begin{eqnarray}
\lambda &=& -\ln \left({4\cos^2(\alpha) - 1 \over 3}\right) \approx {4 \alpha^2 \over 3} + O(\alpha^4)
\label{lambda}
\end{eqnarray}
and $f_0^{(j)} = \Tr_j[\rho_0^{(j)}]$. To be precise, a real $\lambda$ like (\ref{lambda}) is valid for $\alpha < \pi/6 \approx 0.52$, 
otherwise $\langle f^{(j)}(t) \rangle$ oscillates. For coherent errors with non-constant distributions
the result is very similar: for each qubit, the $\lambda$ in (\ref{f^j_coherent}) must be replaced by the respective $\lambda_j$.

We also have closed expressions for the incoherent scenarios proposed, including the additional averages over the Gaussian distribution of coefficients and
already assuming $\alpha, \sigma_{\alpha} << 1$. For incoherent noise with long correlation time:
\begin{eqnarray}
\langle f^{(j)}(t) \rangle \approx {1 \over 2} + \left(f^{(j)}_0 - {1 \over 2} \right) {\exp(-a(t)) \over \sqrt{1+ {8 \over 3} \sigma^2_{\alpha} t} }
\label{fid1_tb3_b0}
\end{eqnarray}
with
\begin{eqnarray}
a(t) = { {4 \over 3} \alpha^2 t \over 1 + {8 \over 3} \sigma_{\alpha}^2 t}
\end{eqnarray}
For incoherent noise with short correlation time:
\begin{eqnarray}
\langle f^{(j)}(t) \rangle \approx {1 \over 2} + \left(f^{(j)}_0 - {1 \over 2} \right) \exp(-\eta t) 
\label{fid1_tc2_b0}
\end{eqnarray}
with
\begin{eqnarray}
\eta=-\ln \left({1 + 2 \cos(2 \alpha) \exp(-2\sigma_{\alpha}^2) \over 3}\right)
\label{eta}
\end{eqnarray}\\

\begin{figure}[h]
\includegraphics[width=0.38\textwidth, angle=270]{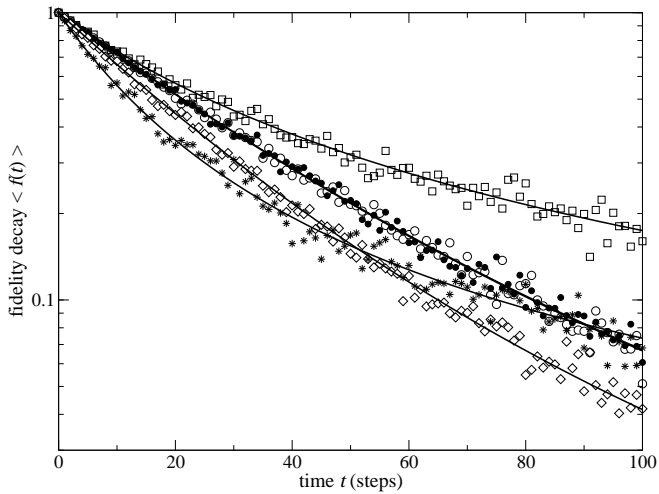}
\caption{\small \label{fig:Zterms_only_curves} Exponential-like decay of $\langle f(t) \rangle $ with only one-body terms in $G_{\tau}$, following our theoretical results (solid lines).
For 8 qubits initially all in the $| 0 \rangle$ state. 
Also we have plotted the numerical calculation using 100 realizations.}
\begin{flushleft}
\small
\vspace{-0.3cm}
$\square$: type IL, $P$ Gaussian with $\alpha=0$, $\sigma_{\alpha}=0.08$. \\
$\bullet$: type IS, $P$ Gaussian with $\alpha=0$, $\sigma_{\alpha}=0.08$. \\
$\Circle$: type C, $P$ constant with $\alpha=0.08$. \\
$\diamondsuit$: type IS, $P$ Gaussian with $\alpha=0.08,\ \sigma_{\alpha}=0.04$. \\
*: type IL, $P$ Gaussian with $\alpha=0.08,\ \sigma_{\alpha}=0.08$. \\
\end{flushleft}
\vspace{-0.3cm}
\begin{tabular*}{8cm}{c}
\hline
\end{tabular*}
\end{figure}

Fig. \ref{fig:Zterms_only_curves} shows some examples of numerical calculations together with the theoretical result, exhibiting perfect agreement.\\

\subsection{Numerical results}

\begin{figure}[h]
\includegraphics[width=0.38\textwidth, angle=270]{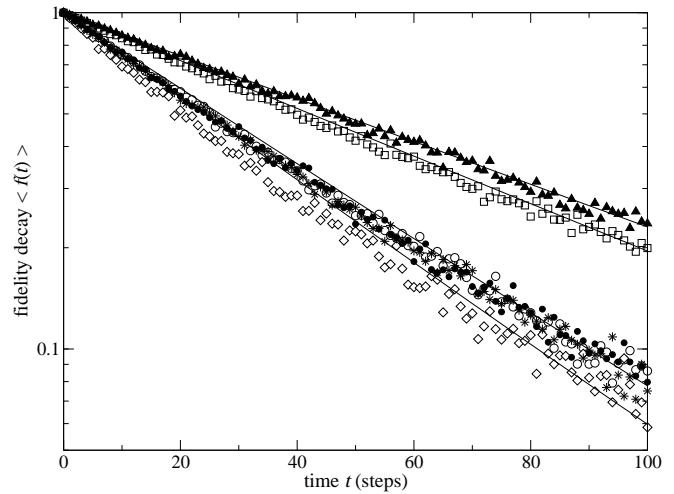}
\caption{\small \label{fig:curves_pure} Examples of the numerical calculation of $\langle f(t) \rangle$:
we examined the different $P$ distributions assuming that only
one-body and first neighbor couplings are present in $G_{\tau}$. For 8 qubits, initially all in the $| 0 \rangle$ state. 
100 realizations.
The fitting corresponds to formula (\ref{Gamma}) with $f_{co} = 0.1$ (see text). }
\vspace{-0.3cm}
\begin{flushleft}
\small
$\blacktriangle$: type C, $P$ constant, $\alpha=0.05,\ \beta=0.02$. \\
$\square$: type C, $P$ constant, $\alpha=0.02,\ \beta=0.05$.\\
$\Circle$: type C, $P$ constant, $\alpha=\beta=0.05$. \\
*: type IS, $P$ Gaussian with $\alpha=\beta=0.05,\ \sigma_{\alpha}=\sigma_{\beta}=0.005$. \\
$\bullet$: type IS, $P$ Gaussian with $\alpha=\beta=0,\ \sigma_{\alpha}=\sigma_{\beta}=0.05$. \\
$\diamondsuit$: type IL, $P$ Gaussian with $\alpha=\beta=0.05, \sigma_{\alpha}=\sigma_{\beta}=0.005$.
\end{flushleft}
\vspace{-0.3cm}
\begin{tabular*}{8cm}{c}
\hline
\end{tabular*}
\end{figure}

\begin{figure}[h]
\includegraphics[width=0.38\textwidth, angle=270]{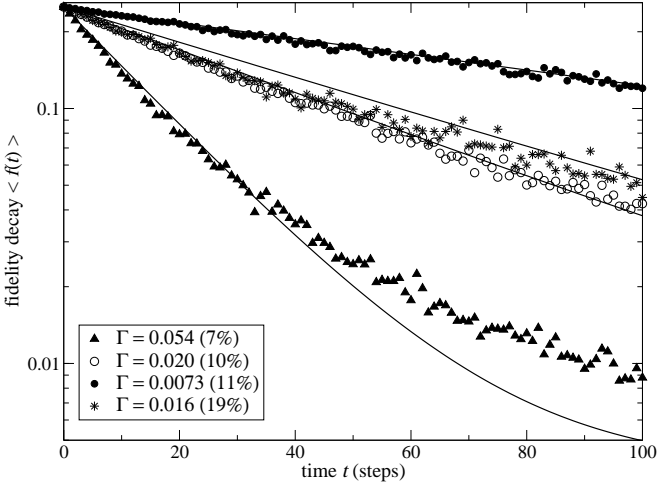}
\caption{\small \label{fig:curves_two_id} Example of the numerical calculation of $\langle f(t) \rangle$: we chose a particular $P$ distribution and we
worked with different noise strengths.
$P$ corresponds to IL noise with coefficients drawn from a Gaussian distribution centered around $0$. 
For 8 qubits; initially, qubits 1 and 2 are in the $\Id/2$ state while the rest is in the $| 0 \rangle$ state. 100 realizations. 
The fitting corresponds to formula (\ref{Gamma}) with $f_{co} = 0.05$ (see text); the inset shows the obtained values of $\Gamma$ with their relative error.}
\vspace{-0.3cm}
\begin{flushleft}
\small
$\bullet$: one-body and first-neighbor terms in $G_{\tau}$, $\sigma_{\alpha}=\sigma_{\beta}=0.03$.\\
*: one-body and first-neighbor terms in $G_{\tau}$, $\sigma_{\alpha}=\sigma_{\beta}=0.05$. \\
$\Circle$: with all the terms in $G_{\tau}$, $\sigma_{\alpha}=\sigma_{\beta}=0.03$. \\
$\blacktriangle$: with all the terms in $G_{\tau}$, $\sigma_{\alpha}=\sigma_{\beta}=0.05$.
\vspace{-0.3cm}
\end{flushleft}
\begin{tabular*}{8cm}{c}
\hline
\end{tabular*}
\end{figure}

When multi-body terms are present in $G_{\tau}$, the non-separability of the $E_t$ prevents us from getting closed results for $\langle f(t) \rangle$. We studied
numerically the case when only one-body and two-body terms are present, obtaining always a linear initial decay and a saturation value for long times, as mentioned in Sec. II.
Figs. \ref{fig:curves_pure} and \ref{fig:curves_two_id} 
show some examples of the numerical calculation of $\langle f(t) \rangle$ and a curve-fitting following the formula 
\begin{equation}
\langle f(t) \rangle = e^{-\Gamma t}\left( f_0 - {1 \over N} \right) + {1 \over N}
\label{Gamma}
\end{equation}
where $\Gamma$ is the only fitting parameter. 
As before, \mbox{$f_0 = \Tr[\rho_0^2]$}.  
We don't expect an exact agreement with this formula; we chose it as it's the simplest exponential-type decay
to interpolate the initial and long time behavior. Also, this choice is motivated by the fact that this is the exact expression
for the fidelity decay when the random operators are rotations on $U(N)$ (c.f. eq. (23) in \cite{noise_estimation_TrE}).

In practice, the fitting is simply a linear fitting of $\log(\langle f(t) \rangle - 1/N)$. 
Of course we must be careful with the values of $\langle f(t) \rangle$ close to $0$ (since in this range numerically we will have null and
negative values of the fidelity); thus we just use the points with $f$ higher than a certain cut-off value $f_{co}$. Notice that for fast decays, $f_{co}$ should be low enough to include
a sufficiently large number of points to fit.


\begin{figure}[h]
\centering
\includegraphics[width=0.38\textwidth, angle=270]{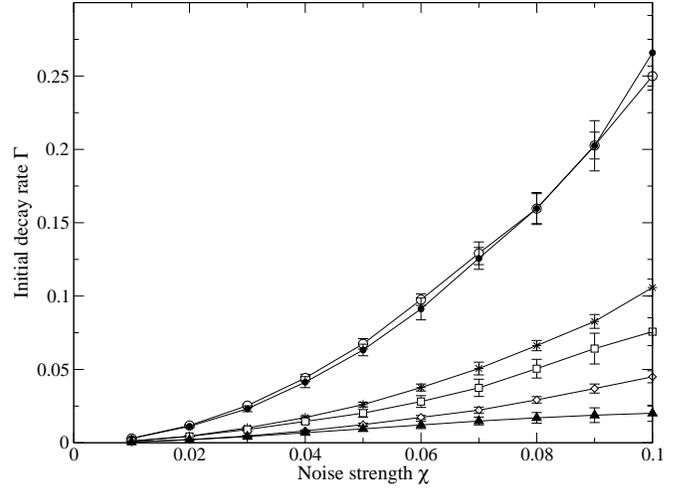}
\caption{\label{fig:Gammas} Values of $\Gamma$ fitting the numerical data according to formula (\ref{Gamma}). We worked with 8 qubits initially all in the $| 0 \rangle$ state.}
\begin{flushleft}
\vspace{-0.3cm}
\small
$\Circle$: type C, $P$ constant with $\alpha=\beta=\chi$; all the terms in $G_{\tau}$. \\
$\bullet$: type IL, $P$ Gaussian with $\alpha=\beta=0$, $\sigma_{\alpha}=\sigma_{\beta}=\chi$; all the terms in $G_{\tau}$. \\
*: idem $\Circle$, but only one-body and first-neighbor terms. \\
$\square$: idem $\bullet$, but only one-body and first-neighbor terms. \\
$\diamondsuit$: idem $\Circle$, but only one-body terms. \\
$\blacktriangle$: idem $\bullet$, but only one-body terms.
\end{flushleft}
\vspace{-0.3cm}
\begin{tabular*}{8cm}{c}
\hline
\end{tabular*}
\end{figure}

Fig. \ref{fig:Gammas} shows the decay rate $\Gamma$ of formula (\ref{Gamma}) as a function of the strength $\chi$, for different noise scenarios. 
It's not surprising to find that the decay is faster for higher $\chi$. Notice also the different proportionality between $\Gamma$
and $\chi$ for the different cases. 
It's worth noticing here that the source for the data $\Gamma \textrm{ vs. } \chi$ in Fig. \ref{fig:Gammas} is the same as for $\gamma \textrm{ vs. } \chi$ in Fig. \ref{fig:gammas}.
The slight
difference lies on the nature of the fitting: in the former, we used eq. (\ref{Gamma}) with $f_{co}=0.1$ while for the later we just used a linear fitting of the points with $f > f_{lim}=0.9$.
The work with $\Gamma$ originally encouraged further studies on the decay rate, while later $\gamma$ became a more fruitful quantity for noise characterization.\\

\section{Appendix C: The initial decay rate $\gamma$}

This appendix is intended to be a short guide to reproduce the formulas (\ref{allpure}), (\ref{gamma_a}), (\ref{gamma_ab}), (\ref{gamma_abc}) and beyond,
giving the reader more general formulae including any multi-body terms and also to account for the effect of a 
non-hermitian $G_{\tau}$ (complex coefficients $\chi_l$, cf. eq. (\ref{generator})). We excluded this level of generality from the main text since the expressions are 
quite cumbersome.

We work on the fidelity at $t=1$, that is after only one iteration of the algorithm, and for a subset of $m$ qubits in $\mathscr{H}^{(M)}$ of the $n$ qubits in $\mathscr{H}_N$ conforming the system under study. These are the $m$ qubits that will be measured (refer to Fig. \ref{fig:circuit_fid_partial}). 
The initial state of the system is $\rho_0^{(M)}\otimes \rho_0^{(\overline{M})}$, where 
$\rho_0^{(M)}= \bigotimes_{j \epsilon M} \rho_0^{(j)}$ is separable but $\rho_0^{(\overline{M})}$ could be any state. 
Following eqs. (\ref{def_fid2}) and (\ref{def_partial_fid}), we want to calculate
\begin{eqnarray}
\langle f^{(M)}(t=1) \rangle &=& \left\langle \Tr_M \left[ \left( R^{(M)} \rho_0^{(M)} R^{\dag (M)} \right)
\left\langle \rho_E^{(M)} \right\rangle_{\overline{M}} \right] \right\rangle_M \nonumber \\
&=& f_0^{(M)}(1 - \gamma^{(M)})
\label{f^M}
\end{eqnarray}
where we have made explicit the average over the random rotations in $M$ and the one over the random rotations in $\overline{M}$.
We have used the notation $\bigotimes_{j \epsilon M} R^{(j)} \equiv R^{(M)}$. $\rho_E^{(M)}$ is
\begin{eqnarray}
\rho_E^{(M)} &=& \Tr_{\overline{M}} \left[ E \left( \bigotimes_{j=1}^n R^{(j)} \right) \rho_0 \left( \bigotimes_{j=1}^n R^{\dag (j)} \right) E^\dag \right] \ \
\label{rhoE}
\end{eqnarray}
where $E=\exp(-i G_{\tau})$, the error operator for this first step, can be expanded as
\begin{eqnarray}
E &=& \Id - i G_{\tau} - \frac{1}{2} G_{\tau}^2 + \frac{i}{6} G_{\tau}^3 + O(\chi^4)
\label{E}
\end{eqnarray}
$G_{\tau}$ is given by eqs. (\ref{generator}) and (\ref{O_l}), as explained in Sec. II, but in principle we will allow for the $\chi_l$ to be complex numbers.
We will rely on the fact that $G_{\tau}$ can be made small so high order powers of $\chi$ will be negligible; this is possible in theory if the time $\tau$ 
can be made arbitrarily small.
We thus insert (\ref{E}) in (\ref{rhoE}) and keep only the terms up to $O(\chi^3)$. Then we use the separability of $\rho_0^{(M)}$, $R^{(M)}$ and the $O_l$ to 
express (\ref{f^M}) as a sum of terms of the form 
 $\langle \Tr[ R^{(j)} \rho_0^{(j)} R^{\dag(j)} O_l^{(j)}  R^{(j)} \rho_0^{(j)} R^{\dag(j)} O_{l'}^{(j)} ] \rangle$,
 $\langle \Tr[ R^{(j)} \rho_0^{(j)} R^{\dag(j)} O_l^{(j)}  R^{(j)} \rho_0^{(j)} R^{\dag(j)} O_{l'}^{(j)} O_{l''}^{(j)}] \rangle$,
 $\langle \Tr[ R^{(j)} \rho_0^{(j)} R^{\dag(j)} O_l^{(j)}] \rangle$, 
 $\langle \Tr[ R^{(j)} \rho_0^{(j)} R^{\dag(j)} O_l^{(j)} O_{l'}^{(j)} O_{l''}^{(j)}] \rangle$ and
 $\langle \Tr[ R^{(j)} \rho_0^{(j)} R^{\dag(j)} O_l^{(j)} O_{l'}^{(j)}] \rangle$.
The average over the random rotations of these quantities can be computed using (\ref{ben1}) -the last three- and (\ref{ben2}) -the first two.
Using the fact that the $O_l^{(j)}$ are either a Pauli matrix or the Identity operator $\Id$, and that $\Tr[\rho_0^{(j)}] = 1$, we arrive 
to a closed expression for (\ref{f^M}) in terms of the purity $\mathcal{P}_j = \Tr_j \left[ \left( \rho_0^{(j)} \right)^2 \right]$ of the initial state of each qubit.
This dependence on the purity only is not suprising; we can picture it the Bloch sphere: the random rotation of the qubits erase the information about the direction
of the polarization vector, but not about its modulus. We obtain:
\begin{eqnarray}
\gamma^{(M)} &=& \sum_{l} Re[ \chi_l ]^2  \left( \prod_{j \epsilon M} \mathcal{P}_j -  \prod_{j \epsilon M} \mathcal{C}_j(l) \right) \nonumber \\
&-& \sum_{l} Im[ \chi_l ]^2  \left( \prod_{j \epsilon M} \mathcal{P}_j + \prod_{j \epsilon M} \mathcal{C}_j(l) \right)
+ O(\chi^3) \hspace{0.7cm}
\label{general_gamma}
\end{eqnarray}
where
\begin{eqnarray}
\mathcal{C}_j(l) = \left\{ \begin{array}{ll} 
          (2/3) (1 - \mathcal{P}_j/2) & \textrm{ if $O_l^{(j)} = $ a Pauli matrix}\\
          \mathcal{P}_j & \textrm{ if $O_l^{(j)}= \Id$} \end{array} \right.
\nonumber
\end{eqnarray}
If the $\chi_l$ are real, not only the second term in (\ref{general_gamma}) vanishes but also it can be proved that the terms of cubic order $O(\chi^3)$ vanish as well.

\end{document}